\documentclass[pra,superscriptaddress,oneside,10pt,twocolumn]{revtex4}
\usepackage{amssymb}
\usepackage{graphicx}

\input{tcilatex}

\begin{document}

\title{Multi-Player and Multi-Choice Quantum Game}
\author{DU Jiang-Feng }
\email{djf@ustc.edu.cn}
\affiliation{Laboratory of Quantum Communication and Quantum Computation, University of
Science and Technology of China, Hefei 230026}
\affiliation{Department of Modern Physics, University of Science and Technology of China,
Hefei 230027}
\author{LI Hui }
\email{lhuy@mail.ustc.edu.cn}
\affiliation{Department of Modern Physics, University of Science and Technology of China,
Hefei 230027}
\author{XU Xiao-Dong}
\affiliation{Department of Modern Physics, University of Science and Technology of China,
Hefei 230027}
\author{ZHOU Xian-Yi }
\affiliation{Laboratory of Quantum Communication and Quantum Computation, University of
Science and Technology of China, Hefei 230026}
\affiliation{Department of Modern Physics, University of Science and Technology of China,
Hefei 230027}
\author{HAN Rong-Dian}
\affiliation{Laboratory of Quantum Communication and Quantum Computation, University of
Science and Technology of China, Hefei 230026}
\affiliation{Department of Modern Physics, University of Science and Technology of China,
Hefei 230027}

\begin{abstract}
We investigate a multi-player and multi-choice quantum game. We start from
two-player and two-choice game and the result is better than its classical
version. Then we extend it to $N$-player and $N$-choice cases. In the
quantum domain, we provide a strategy with which players can always avoid
the worst outcome. Also, by changing the value of the parameter of the
initial state, the probabilities for players to obtain the best payoff will
be much higher that in its classical version.
\end{abstract}

\maketitle

Compared to the long history of mathematics and human conflict, game theory
is a relatively recent creation. Since the first book of Von Neumann and
Oskar Morgenstern,\cite{book} game theory has shed new light on economics,
social science and evolutionary biology.\cite{add3} In view of the
information hold by the players, games can be classified into different
types. An important type of games is \textit{complete information static
games}, in which the players cannot obtain any information about actions of
their opponents and cannot communicate with each other. In this kind of
games, each player can only select his/her strategy depending on information
of his/her own. Therefore the result of the game may not be the best
possible one. A famous example is the traditional \textit{Prisoner's Dilemma}%
. Another reason that contributes to this situation is that each rational
player would like to maximize his/her individual advantage rather than the
collective benefit of all the players (this is actually a key assumption in
both the classical game theory and the quantum game theory).

Quantum game theory is a new born branch of quantum information theory. In a
quantum game, even though the players cannot obtain any information about
actions of their opponents (such as in classical ones), but things are
totally different due to the possible entanglement between the states of the
players. When states of the players are entangled, local operation of a
single player on his/her own state can affect states of other players, so
the payoff for the players could be different from (maybe better than) that
in the classical version of the game. This novel feature led the game theory
into the quantum domain and some marvelous results were found.\cite{6,7,8}
Goldenberg \textit{et al.}\cite{1} presented a two-party protocol for
quantum gambling and their protocol allows two remote parties to play a
gambling game, such that in a certain limit it becomes a fair game. Meyer%
\cite{2} investigated the PQ Game --- a coin tossing game --- and found out
that one player could increase his payoff by implementing quantum strategy
against his classical opponent. Eisert \textit{et al.}\cite{3} quantized the
Prisoner's Dilemma. For this particular case they showed that the game
ceases to pose a dilemma if quantum strategies are allowed for. Marinatto 
\textit{et al.} investigated the quantization of the \textit{Battle of Sexes
Game}.\cite{pla1} Research on evolutional procedure of quantum games are
also presented.\cite{pra1,pla2} Benjamin \textit{et al.}\cite{4,newsud}
presented the study of quantum games with more than two players. They
demonstrated that such games can exhibit "coherent" equilibrium strategies.
All these works were based on maximally entangled states. Recently, we
investigated the correlations between entanglement and quantum games.\cite%
{our-pla} The results showed that for the particular case of the quantum
Prisoners' Dilemma, the property of the game changes fascinatingly with the
variation of the measure of the game entanglement. What is more, although
quantum games are mostly explored theoretically, we have just successfully
realized the quantum Prisoners' Dilemma experimentally on a nuclear magnetic
resonance (NMR) quantum computer.\cite{our-prl,newsud}

Two things have been taken into consideration by all the researchers. The
first is that the players care most about the expected payoff, which is the
average payoff the players can obtain after the game being repeated many
times. The second is that each of the players has only ``a qubit,'' which
means that each of them will end in two states, either $\left|
0\right\rangle $ or $\left| 1\right\rangle $, after the outcome state being
measured. Yet in fact we can find many cases in which these two points are
not of vital importance. Furthermore, cases that do not contain these two
points can also be found. In many cases, the game can not be played for the
second time. In these cases, the players care most about what payoffs they
can at least obtain. Thus each of them will take the best to avoid the worst
outcome. Here the payoff they can at least obtain becomes more important to
the players than the average payoff. In addition, in many cases, the players
have more than two choices. When the players have $k$ choices (denoted by $%
\left| 0\right\rangle ,\left| 1\right\rangle ,\cdots ,\left|
k-1\right\rangle $), the state of each player should be superposition of $%
\left| 0\right\rangle ,\left| 1\right\rangle ,\cdots ,\left|
k-1\right\rangle $. The operation of a single player can be represented by a 
$k$-dimension unitary operator.

In this Letter, we investigate multi-player and multi-choice games by
quantizing the ``Truckers Game,''\cite{5}(specified as followed) which is
significant in social life. We start by investigating a game of two
truckers. Then we generalize it to the $N$-trucker game. The result shows
that if the truckers play in the quantum world, they can always avoid the
worst outcome while in the classical world the worst outcome will always
occur with certain probability. The probability of the best outcome can be
much higher than that in the classical world if a different value is set for
the parameter in the initial state.

We suppose that $N$ truckers are planning to go to city B from city A. There
are $N$ roads connecting the two cities and all of them are assumed to have
the same length. Since each of the $N$ truckers does not know other
truckers' choices, so each of them can only choose his way randomly. The
payoff for a certain trucker depends on how many truckers choose the same
road as he/she does. The more truckers choose the same road, the less payoff
this certain trucker obtains. If all of them choose the same road, the
outcome is the worst because of the possible traffic jamming. The
probability of this situation is $P_{worst}^{C}$. 
\begin{equation}
P_{worst}^{C}=\frac{N}{N^{N}}  \label{eq 1}
\end{equation}%
The superscript $C$ denotes classical condition. On the contrary, if they
all choose different roads, the outcome is the best because there will not
be any traffic jamming. The probability of this situation is $P_{best}^{C}$. 
\begin{equation}
P_{best}^{C}=\frac{N!}{N^{N}}  \label{eq 2}
\end{equation}

It is obvious that in order to avoid the worst outcome, the truckers will do
their best to avoid choosing the same road. Since they can not obtain
information from each other, the worst outcome will happen with certain
probability $P_{worst}^{C}=N/N^{N}$. If we quantize this game, the truckers
can definitely avoid the worst outcome by implementing quantum strategies
(without knowing what other truckers choose) and the probability to obtain
the best outcome can be much higher than that in classical game if a
different value of the parameter in the initial state is set.

\begin{figure}[tbp]
\includegraphics{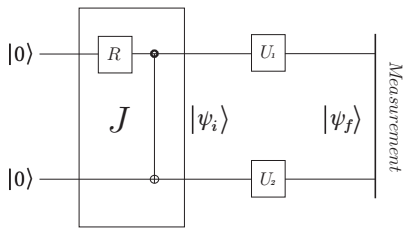}
\caption{The setup of the two-trucker game.}
\label{Figure1}
\end{figure}

We first present the simple two-trucker game. The physical model for this
situation is given in figure(\ref{Figure1}).

We send each player a $2$-state system in the zero state. The input state is 
$\left| \psi \right\rangle =\left| 00\right\rangle $. The gate $R$ is $R=%
\frac{1}{\sqrt{2}}\left( 
\begin{array}{cc}
1 & 1 \\ 
-1 & 1%
\end{array}%
\right) $. Here $U_{1}$ and $U_{2}$ are the strategies that two truckers
adopt respectively.

The state after gate $J$ is $\left| \psi _{i}\right\rangle =\frac{1}{\sqrt{2}%
}\left( \left| 00\right\rangle -\left| 11\right\rangle \right) $. If $%
U_{1}=U_{2}=H$, here $H=\frac{1}{\sqrt{2}}\left( 
\begin{array}{cc}
1 & 1 \\ 
1 & -1%
\end{array}%
\right) $ is the Hadamard gate, the final state is $\left| \psi
_{f}\right\rangle =\frac{1}{\sqrt{2}}\left( \left| 01\right\rangle +\left|
10\right\rangle \right) $. From $\left| \psi _{f}\right\rangle $, we can see
that the two truckers are definitely in different roads. Thus, the
probability of the best outcome is $1$ and the worst situation is avoided.
This is the result that the truckers want.

After we investigated quantum two-trucker game, now we study the general
version of this game which includes $N$ truckers and $N$ roads.

We number the truckers from $0$ to $N-1$ and the roads from $0$ to $N-1$. If
trucker $0$ chooses the road $j_{0}$, trucker $1$ chooses the road $j_{1}$%
,..., and trucker $N-1$ chooses the road $j_{N-1}$, then the state is
described by $\left| j_{0}j_{1}\cdots j_{N-1}\right\rangle $. The game is
started from the state $\left| 00\cdots 0\right\rangle $. Then a
transforming gate is used to obtain the initial state which is denoted by $%
\left| \psi _{i}\right\rangle $, see equation(\ref{eq 3}). 
\begin{equation}
\left| \psi _{i}\right\rangle =\frac{1}{\sqrt{N}}\sum\limits_{k=0}^{N-1}%
\omega _{N}^{k\cdot p}\left| kk\cdots k\right\rangle  \label{eq 3}
\end{equation}%
where $\omega _{N}=e^{2\pi i/N}$. Here $p$ is the parameter that determines
the phases of the terms in the expression of the initial state. It is
obvious that $\left| \psi _{i}\right\rangle $ is symmetric with respect to
the interchange of the truckers. Now the truckers make their decisions on
which way to go. We assume that all of them adopt the same strategy. The
rationality partially comes from the game symmetry. However the symmetry of
the game does not guarantee that all the players choose the same strategy.
Even symmetric games can have asymmetric solutions.\cite{our-prl} While in
practice, if a symmetric game has a solution consisting of different
strategies, the players would be confused in choosing one from them.
Therefore a symmetric solution will be more advantageous than an asymmetric
one. Asymmetric solutions will be impracticable in symmetric games.

We denote the strategic operator by $U$. It is obvious that $U$ is an $N$%
-dimension unitary operator that performs on the state of an individual
trucker. The explicit expression of $U$ is 
\begin{equation}
U=\left( u_{ij}\right) _{N\times N}\text{ , }u_{ij}=\frac{1}{\sqrt{N}}%
(\omega _{N})^{i\cdot j}  \label{eq 4}
\end{equation}%
where $i,j=0,1,\cdots N-1$, and $\omega _{N}=e^{2\pi i/N}$. $U$\ is unitary
because 
\begin{equation}
\sum\limits_{k=0}^{N-1}u_{ki}^{\ast }u_{kj}=\frac{1}{N}\sum%
\limits_{k=0}^{N-1}\omega _{N}^{k(j-i)}=\delta _{ij}\text{ , }U^{+}U=I
\label{eq 6}
\end{equation}

As the trucker operators are denoted by $U$, so $\left| \psi
_{i}\right\rangle $ is performed by $U^{\otimes N}$. The final state $\left|
\psi _{f}\right\rangle $ is%
\begin{eqnarray}
\left| \psi _{f}\right\rangle &=&U^{\otimes N}\left| \psi _{i}\right\rangle 
\nonumber \\
&=&\left( \frac{1}{\sqrt{N}}\right)
^{N+1}\sum\limits_{k=0}^{N-1}\sum\limits_{j_{0}=0}^{N-1}\cdots
\sum\limits_{j_{N-1}=0}^{N-1}  \nonumber \\
&&\left( \omega _{N}^{k\cdot p}u_{kj_{0}}\cdots u_{kj_{N-1}}\left|
j_{0}\cdots j_{N-1}\right\rangle \right)  \label{eq 7}
\end{eqnarray}%
Therefore the coefficient of $\left| j_{0}\cdots j_{N-1}\right\rangle $ is
apparent 
\begin{eqnarray}
C_{j_{0}\cdots j_{N-1}} &=&\left( \frac{1}{\sqrt{N}}\right)
^{N+1}\sum\limits_{k=0}^{N-1}\omega _{N}^{k\cdot p}u_{kj_{0}}\cdots
u_{kj_{N-1}}  \nonumber \\
&=&\left( \frac{1}{\sqrt{N}}\right) ^{N+1}\sum\limits_{k=0}^{N-1}\omega
_{N}^{k\cdot m}  \label{eq 8}
\end{eqnarray}%
where $m=j_{0}+\cdots +j_{N-1}+p$.

In some cases the truckers want to guarantee that the payoffs they can at
least obtain is better than the one when they all choose the same way, the
parameter $p$ can here be set as $p=1$. Thus $\left\vert \psi
_{i}\right\rangle $ is 
\begin{equation}
\left\vert \psi _{i}\right\rangle =\frac{1}{\sqrt{N}}\sum\limits_{k=0}^{N-1}%
\omega _{N}^{k\cdot 1}\left\vert kk\cdots k\right\rangle   \label{eq 9}
\end{equation}%
Therefore 
\begin{eqnarray}
C_{j_{0}\cdots j_{N-1}} &=&\left( \frac{1}{\sqrt{N}}\right)
^{N+1}\sum\limits_{k=0}^{N-1}\omega _{N}^{k}u_{kj_{0}}\cdots u_{kj_{N-1}} 
\nonumber \\
&=&\left( \frac{1}{\sqrt{N}}\right) ^{N+1}\sum\limits_{k=0}^{N-1}\omega
_{N}^{k\cdot m}  \label{eq 10}
\end{eqnarray}%
where $m=j_{0}+\cdots +j_{N-1}+1$. When $j_{0}=j_{1}=\cdots =j_{N-1}=j$, the
outcome is the worst and the coefficient of $\left\vert jj\cdots
j\right\rangle $ is 
\begin{equation}
C_{jj\cdots j}=\left( \frac{1}{\sqrt{N}}\right)
^{N+1}\sum\limits_{k=0}^{N-1}\omega _{N}^{k\cdot (jN+1)}=0  \label{eq 11}
\end{equation}%
From equation(\ref{eq 11}) we see that the worst situation will not occur.
The payoffs the truckers can at least obtain can be definitely better than
that of the worst outcome.

If the truckers want to increase the probability of the occurrence of the
best outcome, the parameter can be set as $p=N(N-1)/2$. Thus $\left| \psi
_{i}\right\rangle $ is 
\begin{equation}
\left| \psi _{i}\right\rangle =\frac{1}{\sqrt{N}}\sum\limits_{k=0}^{N-1}%
\omega _{N}^{k\cdot N(N-1)/2}\left| kk\cdots k\right\rangle  \label{eq 12}
\end{equation}%
When $j_{0},j_{1},\cdots ,j_{N-1}$ are different from each other, there is
only one truck in one road and the situation is the best. The probability of
this case is 
\begin{equation}
P_{best}^{Q}=N!\cdot \left| C_{01\cdots N-1}\right| ^{2}=N\cdot N!/N^{N}
\label{eq 13}
\end{equation}%
The superscript Q denotes quantum condition. Compared with the classical
probability given in Eq. (\ref{eq 2}), we have 
\begin{equation}
P_{best}^{Q}=N\cdot P_{best}^{C}  \label{eq 14}
\end{equation}%
Thus the quantum probability is $N$ times higher than the classical one.

From the above description, we see that the outcome state of the game is
closely related to the value of parameter $p$ in the initial state. By
changing the value of parameter $p$, the various needs of the truckers can
be catered to. If the truckers expect a higher probability of the best
outcome, the value of parameter $p$ can be set as $\frac{N(N-1)}{2}$. On the
other hand, if they demand the removal of the worst situation, the value of
parameter $p$ can be set as $1$.

To conclude, we suggest that, besides the expected payoff, the payoff that a
player can at least obtain in an unrepeatable game is also important and
worth commenting. In most common cases, the players tend to have several
choices instead of only two. From this perspective, we investigate the game
of the truckers, in which each trucker can only play for one time and has
more than two roads to choose from. Assuming that all the players adopt the
same strategy, then the game is symmetric with respect to the interchange of
the players, the truckers can always avoid choosing the same way. While in
the classical version of this game, the worst outcome will occur with a
certain probability. If the truckers care most about the probability for the
best outcome to occur, in which all truckers choose different roads, the
value of the parameter $p$ can be reset to make the probability much higher
than that in the classical game.

Due to the complexity of this N-player and N-choice game, we just present
the novel features that accompany the worst and the best outcome of the
game. Since the coefficient described in Eq. (\ref{eq 8}) is either $\sqrt{%
N/N^{N}}$ or $0$, and the square norm of the coefficient is the probability
that the final state (after being measured) collapses into the corresponding
basis, so the probability of the result is $N/N^{N}$, which is $N$ times
higher than that in the classical game, or just $0$. Whether the probability
is $0$ or not depends on both the basis and the initial state of the game.
Different from the classical game, in the quantum game by setting different
values of parameter $p$, truckers can always meet their various needs,
removing the worst outcome or increasing the probability for the best
outcome to occur, without knowing the choices of other truckers.

*Supported by the National Nature Science Foundation of China under Grant
Nos 10075041 and 10075044, and the Science Foundation for Young Scientists
of USTC.

\end{document}